\documentclass[pra,floatfix,showpacs,twocolumn,tightenlines,nofootinbib,superscriptaddress]{revtex4-1}

\usepackage{amsmath}
\usepackage{amsfonts}
\usepackage{amssymb}
\usepackage{graphicx}
\usepackage{braket}
\newcommand{\dg}{^{\dagger}}
\usepackage{lipsum}
\usepackage{fullpage}
\usepackage{appendix}
\usepackage{color}

\newcommand{\hc}{\text{H.c.}}
\newcommand{\dd}{\mathrm{d}}

\begin{document}

\title{Quantum simulation of Unruh-DeWitt detectors with nonlinear optics}

\author{Eugene Adjei}
\affiliation{Perimeter Institute for Theoretical Physics, Waterloo, ON,  Canada, N2L 2Y5}
\affiliation{Department of Physics \& Astronomy, University of Waterloo, Waterloo, ON, Canada, N2L 3G1}
\author{Kevin J. Resch}
\affiliation{Department of Physics \& Astronomy, University of Waterloo, Waterloo, ON, Canada, N2L 3G1}
\affiliation{Institute for Quantum Computing, University of Waterloo, Waterloo, ON, Canada, N2L 3G1}
\author{Agata M. Bra\'nczyk}
\affiliation{Perimeter Institute for Theoretical Physics, Waterloo, ON,  Canada, N2L 2Y5}

\begin{abstract}
We propose a method for simulating an Unruh-DeWitt detector, coupled to a 1+1-dimensional massless scalar field, with a suitably-engineered $\chi^{(2)}$ nonlinear interaction. In this simulation, the parameter playing the role of the detector acceleration is played by the relative inverse-group-velocity gradient inside the nonlinear material. We identify  experimental parameters that tune  the detector energy gap, acceleration, and switching function.  This system can simulate time-dependent acceleration, time-dependent detector energy gaps, and non-vacuum initial detector-field states. Furthermore,  for very short materials, the system can simulate the weak anti-Unruh effect, in which the response of the detector decreases with acceleration.
While some Unruh-related phenomena have been investigated in nonlinear optics, this is the first proposal for simulating an Unruh-DeWitt detector in these systems.
\end{abstract}

\maketitle

The Unruh-DeWitt (UDW) detector model \cite{unruh1976notes,dewitt1979general,unruh1984happens} predicts that an accelerating observer in vacuum will see blackbody radiation where an inertial observer would see none---this  is known as the \emph{Unruh effect} \cite{earman2011unruh}. Due to the prohibitively large accelerations required, however, the effect is yet to be verified experimentally. Nevertheless, one can simulate the physics of this effect by engineering a Hamiltonian of the same form in a different, more controllable, system. Experimental realization of a quantum simulation of the Unruh effect could  generate new insight and stimulate ideas across different fields of physics.

The Unruh effect  is  closely connected to the phenomenon of two-mode squeezing \cite{walls2007quantum}, well-known in quantum nonlinear optics (NLO). This makes NLO a natural candidate for simulating the effect, and indeed some related phenomena have already been studied---specifically, simulations of  black-hole horizons \cite{philbin2008fiber} and the Unruh-Davies effect \cite{guedes2019spectra}. Several other systems have also been investigated for simulating physics related to the Unruh effect \cite{fedichev2003gibbons,fedichev2004observer, paraoanu2014recent,rodriguez2017synthetic,hu2019quantum}, but NLO holds particular promise due to our ability to fabricate and control such systems. New aspects of Unruh-effect-related phenomena are still being discovered, e.g. \cite{carballo2019unruh}, and their simulation in NLO could produce new insights.  

\begin{figure}[t!]
    \centering
        \includegraphics[width=\columnwidth]{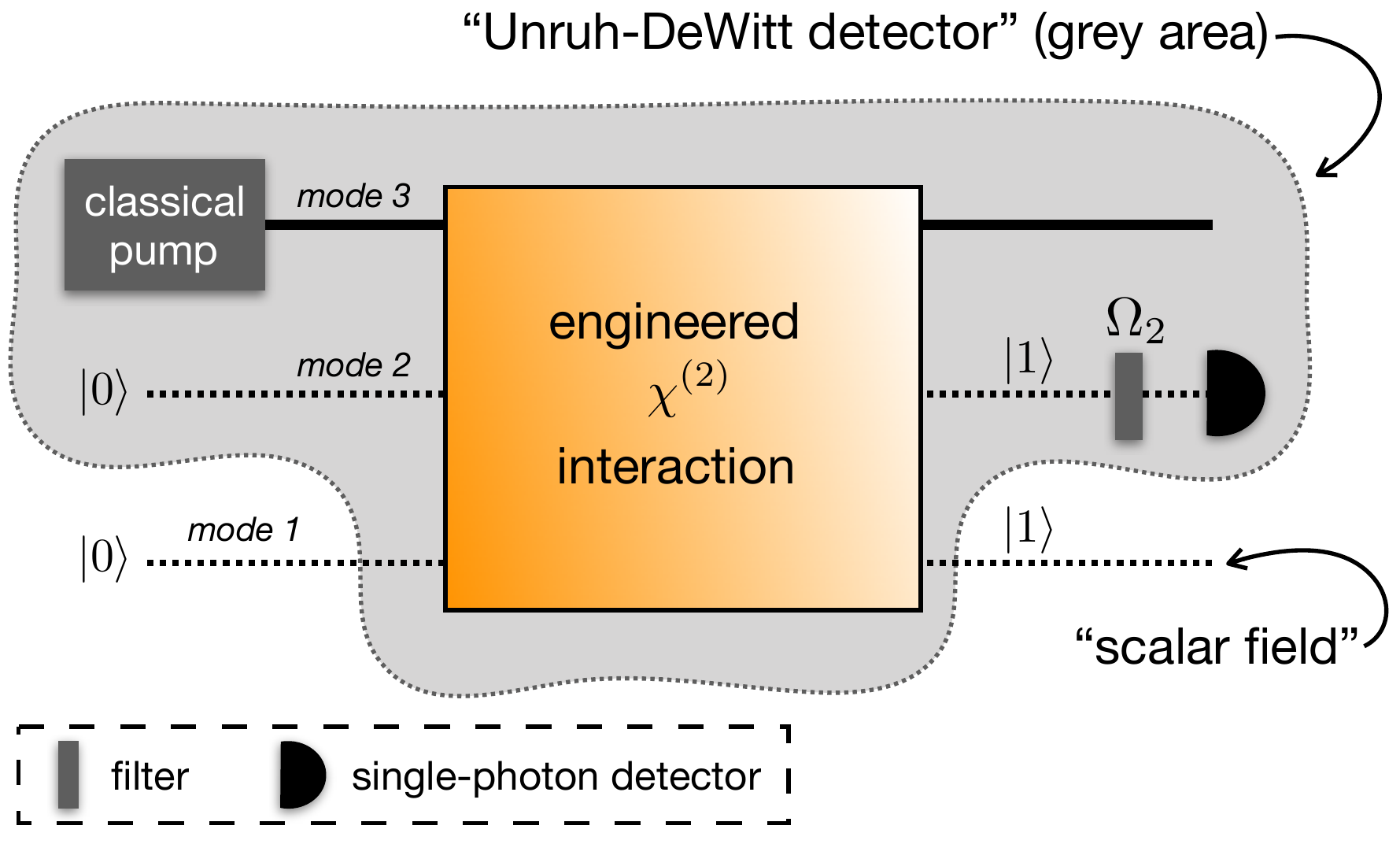}
    \caption{Schematic representing three modes interacting via an engineered $\chi^{(2)}$ interaction (input on the left and output on the right). Modes 1 and 2 are initially prepared in the vacuum state $\ket{0}$, while mode 3 contains a classical pump field. During interaction, a pair of photons is created (one photon in mode 1 and the other in mode 2). An analogy is made between  the $\chi^{(2)}$ interaction (Eq. \eqref{eq:H2b}) and a UDW detector coupled to a massless scalar field (Eq. \eqref{eq:HUD}): mode 1 maps to the scalar field while mode 2 (along with other elements in the shaded grey region) maps to the UDW detector. The shaded grey area encloses the elements that determine the UDW detector energy gap, i.e., the frequency of the pump field, the properties of the nonlinear material, and the filter frequency $\Omega_2$,  as described in Table \ref{tab:table1}. Detection of a photon in mode 2 corresponds to UDW detection.}
    \label{fig:fig1}
\end{figure}

In this paper, we consider a quantum simulation of the Unruh effect formulated in terms of UDW detectors. Concretely, we make the analogy between the interaction of two quantum modes of a downconversion process (FIG. \ref{fig:fig1}) and the interaction  of a non-inertial detector---i.e. a UDW detector---with a scalar field. { This is in contrast to what has been done before where the downconverted state was the analogue for the TMSV of the Rindler wedges \cite{alsing2004teleportation}. } We compute the first-order transition probability amplitudes for both processes and compare them. In doing so, we draw an analogy between  the relative inverse-group-velocity gradient inside the nonlinear material and the acceleration of the UDW detector.

Our proposal explicitly identifies which experimental parameters play the role analogous to the detector energy gap, acceleration, and switching function.  The downconversion can  simulate  time-dependent acceleration, time-dependent detector energy gaps \cite{olson2011entanglement}, and non-vacuum initial detector-field states \cite{aspachs2010optimal}. Furthermore, we show that for very short crystals, one can also simulate the anti-Unruh effect \cite{brenna2016anti}, specifically, the weak anti-Unruh effect \cite{garay2016thermalization} in which the response of the detector decreases with acceleration. 
garay2016thermalization
\section{The Unruh-DeWitt detector}

An Unruh-DeWitt detector couples to a 1+1-dimensional  scalar, massless Klein-Gordon  field $\phi(x(\tau),t(\tau))$ through the  Hamiltonian \cite{garay2016thermalization, scully1997}:
\begin{align}\label{eq:HUD}
H(\tau)={}&\lambda\eta(\tau) m(\tau)\phi(x(\tau),t(\tau))\,,
\end{align}
where $\lambda$ is a small coupling constant with dimension length$^{-1}$, $\eta(\tau)$ is a dimensionless switching function, and $m(\tau)$ is the detector's monopole operator (dimensionless). Here, and in the rest of this paper, we  work with natural units $\hbar=c=1$. The time evolution of $m(\tau)$, where $\tau$ is the proper time in the frame of the detector, is assumed to be $m(\tau)=e^{iH_0\tau}m(0)e^{-iH_0\tau}$, where $H_0$ is the free Hamiltonian of the detector. The coordinates $t$ and $x$ describe the Minkowski coordinates associated to an inertial frame.  

The scalar field $\phi(x(\tau), t(\tau))$ can be expanded in plane-wave solutions of the Klein-Gordon equation. For massless fields propagating in a single direction it is conventional to work with frequency $\omega_k$ rather than wavenumber labels. We thus expand the field (in the interaction picture) in terms of plane waves:
\begin{equation}\label{dimen}
\phi(x(\tau), t(\tau)) = \int \frac{ d\omega_{k}}{\sqrt{4\pi\omega_{k}}} (a(\omega_{k}) e^{-i\omega_{k}q(\tau)} +\mathrm{h.c.})\,,
\end{equation}
where $q(\tau)= t(\tau)-x(\tau)$ and $[a(\omega_{k}), a^{\dagger}(\omega_{k'})] = \delta(\omega_{k} - \omega_{k'})$. As  evident from Eq. (\ref{dimen}), the scalar field is dimensionless (in natural units), as expected for a field  defined on 1+1 spacetime \cite{milonni2013quantum}. 

Our goal is to show that the UDW Hamiltonian in Eq. \eqref{eq:HUD} can be simulated with a quantum NLO Hamiltonian. Modifying the quantum NLO Hamiltonian directly to account for the analogue of detector acceleration poses  significant technical challenges. In this paper, we thus consider a more manageable approach, and show the equivalence between the two systems by comparing the transition amplitudes at first order. We leave demonstrating the equivalence between the Hamiltonians for future work.  

We start the detector in the ground state $\ket{g}$, and take the field to be in the Minkowski vacuum state $\ket{0}$. For a detector turned on at time $\tau_i$ and off at time $\tau_f$, the conditional probability amplitude  (to first order), $\mathcal{A}$, of finding the detector in the excited state $\ket{e}$ given the field is in the single-particle state $\ket{\omega}=a\dg(\omega)\ket{0}$, is
\begin{align}\label{eq:AUDx}
   \mathcal{A} ={}&-i \bra{\omega}\bra{e}\int_{\tau_i}^{\tau_f} \dd \tau H(\tau)\ket{0}\ket{g}\,.
\end{align}
In Appendix \ref{sec:UDWA}, we show that this can be rewritten as
\begin{align}\label{eq:AUDx2}
   \mathcal{A} ={}&b(\omega)\int_{\tau_i}^{\tau_f} \dd \tau \eta(\tau)e^{i\Omega\tau} e^{i\omega q(\tau) }\,,
\end{align}
with $ b(\omega)=-i\lambda /\sqrt{4\pi\omega} \bra{e}m(0)\ket{g}$  and where $\Omega=E_e-E_g$ is the energy gap of the detector. The transition amplitude is not typically written this way, so to relate the expression in Eq. \eqref{eq:AUDx2} to known results, we briefly discuss a few special cases. 

For an inertial detector, $t(\tau)= \gamma \tau $ and  $x(\tau)=x_0+vt=x_0+v\gamma\tau$ where $\gamma=(1-v^2)^{-1/2}$  and $\tau_f=-\tau_i=T$, and thus $   \mathcal{A} =e^{ikx_0} \mathrm{sinc}((\Omega+\gamma(\omega-kv))T)/T\sqrt{\pi\omega}$, as found in    \cite{sriramkumar1996finite}. In the limit $T\rightarrow\infty$, one gets $\mathcal{A} \sim\delta(\Omega+\gamma(\omega-kv))$. Since $kv\leq|k||v|<\omega$ and $\Omega>0$, the argument of the delta function is always positive and the detection is forbidden on the grounds of energy conservation. For a uniformly accelerating observer, $x(\tau)= a^{-1}\cosh(a\tau)$ and $t(\tau)= a^{-1}\sinh(a\tau)$, giving $  q(\tau)=-a^{-1}e^{-a\tau}$. The limit $T\rightarrow\infty$, yields the famous result that an observer, accelerating through a vacuum, sees a thermal field. Concretely, in this limit,  $ |\mathcal{A}|^2=(e^{2\pi\Omega /a}-1)^{-1}/2a\Omega\omega$, which has a Planckian form in $\Omega$ with a temperature $\beta^{-1}=a/2\pi$ \cite{sriramkumar1996finite}.

\section{The analogy with SPDC}In a $\chi^{(2)}$ nonlinear-optical process, a nonlinear material mediates the interaction between three photons. An example of this is SPDC, in which high-energy pump photons are converted into pairs of lower energy photons. $\chi^{(2)}$ processes have widespread application in quantum computation \cite{Kok2007}, quantum communication \cite{Gisin2007} as sources of non-classical light and quantum interactions \cite{donohue2015theory}, and quantum metrology \cite{Higgins2007,Nagata2007}, as well as in more specialized areas such as quantum imaging \cite{Brida2010}, quantum lithography \cite{Boto2000}, or optical coherence tomography \cite{Nasr2008}. Here, we consider a new application as an analogue of the UDW detector.

The interaction Hamiltonian for a three-wave mixing process in a waveguide is \cite{yang2008spontaneous}:
\begin{align}\label{eq:H2b}
\begin{split}
H_\textsc{s}(t)={}&-\frac{1}{3\epsilon_0}\int d\mathbf{r}\Gamma_{ijk}^{(2)}(\mathbf{r})\hat{D}^{i}(\mathbf{r},t)\hat{D}^{j}(\mathbf{r},t)\hat{D}^{k}(\mathbf{r},t)\,,
\end{split}
\end{align}
where $\Gamma^{(2)}(\mathbf{r})$ is a tensor related to the more commonly-used nonlinear tensor $\chi^{(2)}(\mathbf{r})$ via Eq. (15) in \cite{yang2008spontaneous}, and $\hat D$ are the quantized displacement field operators.

Let us assume that we have three fields present in the process (labelled by $1,2,3$), which all have zero overlap with each other, either due to non-overlapping frequencies or orthogonal polarization. Let us also assume that one of the fields ($3$) is intense, non-depleting, and can be treated classically. Under this assumption, both Eqs. (\ref{eq:HUD}) and (\ref{eq:H2b}) describe the interaction between two quantum systems: Eq. (\ref{eq:HUD}) describes the interaction between a UDW detector and a quantized scalar massless field while Eq. (\ref{eq:H2b}) describes the interaction between one quantized EM field and a second quantized EM field.

We now identify the massless scalar field with the field in mode 1, and the UDW detector with the field in mode 2. With this analogy, we  start modes 1 and 2 in the vacuum state and interact them, with the classical pump field centred at $\Omega_3$, in the nonlinear material. After the interaction, we send mode 2 (the UDW detector) through a spectral filter at $\Omega_2$ to fix the UDW detector's energy gap. Subsequent detection of the photon in mode 2 corresponds to UDW detection. This process is shown schematically in FIG. \ref{fig:fig1}.

\begin{table*}
\begin{ruledtabular}
\begin{tabular}{ccc}
UDW parameters&SPDC parameters& Correspondence\\
\hline\hline
Detector trajectory $q(\tau)= t(\tau){-}x(\tau)$ & Relative inverse GV $\Delta v_g(z)^{-1}$ & $ q(\tau)\leftrightarrow\int_{v\tau_i}^{v\tau}\Delta v_g(\zeta)^{-1} d\zeta$\\\hline
Switching function $\eta(\tau)$ & Nonlinear susceptibility $\chi^{(2)}(z)$ & $\eta(\tau)\leftrightarrow\frac{\chi^{(2)}(v\tau)v}{\sqrt{n_1(\Omega_3-\Omega_2;v\tau)n_2(\Omega_2;v\tau)n_3(\Omega_3;v\tau)}}$\\
&Refractive index $n_j(\omega,z)$&\\\hline
Detector energy gap $\Omega(\tau)$ & Phase mismatch $\Delta k_0(z)$ & $  \Omega(\tau)\leftrightarrow(\Delta k_0(v\tau){-}\Delta v_g(v\tau)^{-1}(\Omega_3{-}\Omega_2))v$\\
&Relative inverse GV $\Delta v_g(z)^{-1}$ &\\\hline
 Scaling coefficient $b(\omega)$ & Spectral pump amplitude $\alpha(\omega)$ & $ b(\omega)\leftrightarrow i\kappa     \alpha(\Omega_2+\omega)$\\
 &Coupling strength $\kappa$&\\\hline
 Interaction time $\Delta \tau=\tau_f-\tau_i$&Waveguide length $L=z_f-z_i$&$v\Delta\tau\leftrightarrow L$
\end{tabular}
\end{ruledtabular}
\caption{\label{tab:table1}Summary of UDW parameters and the corresponding SPDC parameters. The detector trajectory $q(\tau)$ determines the acceleration. The phase mismatch $\Delta k_0(z)$ and relative inverse group velocity (GV) $\Delta v_g(z)^{-1}$ are defined below  Eqs.  \eqref{eq:paramS}. The coupling strength $\kappa$ is defined in Appendix \ref{sec:spdcder}. The scaling velocity $v$ is defined above Eq. \eqref{eq:AUDx3}. $\Omega_3$ is the mean pump frequency and $\Omega_2$ is the frequency of the filter in mode 2.}
\end{table*}

To compute the corresponding amplitude, we make some further simplifying assumptions. Let us assume that the fields propagate collinearly along $z$ through a waveguide with only one transverse mode that is uniform over a cross-sectional area. Under the assumptions that the tensor nature of $\chi^{(2)}(\mathbf{r})$ and the vector nature of the mode profiles of the displacement field can be neglected, $\chi^{(2)}(z)$ can be taken to characterize the strength of the nonlinearity along the waveguide. We also consider that the refractive index can vary along the length of the waveguide.

For a waveguide starting at position $z_i$ and ending at position $z_f$, the probability amplitude (to first order), $\mathcal{A}_\textsc{s}$, of finding  mode 1 in a single-particle state $\ket{\omega}=\hat{a}^{\dagger}_1(\omega)\ket{0}$ and mode 2 in a single-particle state $\ket{\Omega_2}=\hat{a}^{\dagger}_2(\Omega_2)\ket{0}$ 
 is
\begin{align}\label{eq:AUDx0}
   \mathcal{A}_\textsc{s} ={}&-\frac{i}{\hbar} \bra{\omega}\bra{\Omega_2}\int_{-\infty}^{\infty} \dd t H_\textsc{s}(t)\ket{0}\ket{0}\,.
\end{align}
In Appendix \ref{sec:spdcder}, we show that this can be written as:
\begin{align}
   \mathcal{A}_\textsc{s} ={}& b_\textsc{s}(\omega)\int_{z_i}^{z_f}dz\tilde\eta(z) e^{i\tilde\Omega(z) z}e^{i\omega\tilde q(z)}\,,
\end{align}
where 
\begin{subequations}\label{eq:paramS}
\begin{align}
   b_\textsc{s}(\omega)={}&i\kappa     \alpha(\Omega_2+\omega)\\
    \tilde\eta(z)={}&\frac{\chi^{(2)}(z)}{\sqrt{n_1(\Omega_3-\Omega_2;z)n_2(\Omega_2;z)n_3(\Omega_3;z)}}  \\
    \tilde \Omega(z)={}&\Delta k_0(z)-\frac{1}{\Delta v_g(z)}(\Omega_3-\Omega_2)\\
    \tilde q(z)={}&\int_{z_i}^{z}\frac{1}{\Delta v_g(\zeta)} d\zeta\,,
\end{align}
\end{subequations}
where $\kappa$ is a constant defined in Appendix \ref{sec:spdcder} $\alpha(\Omega_2+\omega)$ is the shape of the classical pulse defined in Appendix \ref{sec:spdcder}, $\Delta k_0(z)=k_3(\Omega_3;z)-k_2(\Omega_2;z)-k_1(\Omega_3-\Omega_2;z)$ is the phase mismatch between the three fields, and $\Delta v_g(z)^{-1}=v_3(z)^{-1}-v_1(z)^{-1}$ is the relative inverse group velocity between fields in modes 1 and 3, where $v_i(z)^{-1}$ are also defined in Appendix \ref{sec:spdcder}.  

We now introduce a scaling velocity $v=z/\tau=L/\Delta\tau$, where $L=z_f-z_i$ and $\Delta\tau=\tau_f-\tau_i$, and write 
\begin{align}\label{eq:AUDx3}
   \mathcal{A}_\textsc{s} ={}&  b_\textsc{s}(\omega)\int_{\tau_i}^{\tau_f} \dd \tau \eta_\textsc{s}(\tau)e^{i\Omega_\textsc{s}(\tau)\tau} e^{i\omega  q_\textsc{s}(\tau) }\,,
\end{align}
where $\eta_\textsc{s}(\tau)\equiv{} \tilde\eta(v\tau)v$, $\Omega_\textsc{s}(\tau)\equiv{}\tilde\Omega(z) v$ and $ q_\textsc{s}(\tau)\equiv{}  \tilde q(v\tau)$. To complete the analogy between the SPDC  system and the UDW-detector-scalar-field system, we identify $b_\textsc{s}$, $\eta_\textsc{s}$, $\Omega_\textsc{s}$ and $q_\textsc{s}$ in Eq. \eqref{eq:AUDx3} respectively with $b$, $\eta$, $\Omega$ and $q$ in  Eq. \eqref{eq:AUDx2}. This is summarized in Table \ref{tab:table1}, which explicitly identifies which experimental parameters correspond to the detector energy gap, the accelerated detector trajectory, and the switching function.

A stationary observer corresponds to a material of constant relative dispersion: $\Delta v_g(z)=v$.  Simulating a constantly accelerating detector  requires a material with a relative inverse group velocity $\Delta v_g(z)^{-1}$ that changes exponentially along the transverse direction $z$, i.e. an \emph{exponential relative-inverse-group-velocity gradient}.  A variable non-exponential $\Delta v_g(z)^{-1}$ would simulate  a detector with a variable acceleration \cite{ostapchuk2012entanglement}.

Simulating a detector with  a constant detector energy gap $\Omega$ requires the relative linear dispersion $\Delta k_0(z)$ to be tailored to compensate the $z$ dependence on the relative-inverse-group-velocity gradient in $\Delta v_g(z)^{-1}$. Without this compensation in $\Delta k_0(z)$, the system would simulate a UDW detector with a variable energy gap \cite{olson2011entanglement}. 

Shaping $\Delta k_0(z)$ and $\Delta v_g(z)$ is done by shaping the refractive indices. This will influence the switching function $\eta(\tau)$. To achieve a desired $\eta(\tau)$, one must shape the second-order nonlinear susceptibility $\chi^{(2)}(z)$, which may be possible using existing nonlinearity shaping methods \cite{graffitti2017pure}.

Furthermore, one can inject quantum light into modes 1 and 2, to simulate  non-vacuum initial detector-field states \cite{aspachs2010optimal}.

Since typical pump pulses have spectral amplitudes $\alpha(\omega)$ with Gaussian or Lorentzian shapes, $b_\textsc{s}(\omega)$ will not have the same frequency dependence as $b(\omega)$. The spectral amplitude $\alpha(\omega)$ can be shaped using standard pulse shaping methods, but only over a finite frequency range. The analogy is therefore limited to within this range. 

To compute numerical results, we make  two simplifying assumptions. First, we assume that the phase mismatch can be shaped  to compensate for the variation in the relative inverse group velocity. We thus introduce $\epsilon(z)$ to simultaneously parametrize the deviation of the phase mismatch from some mean value $\bar{\Delta k_0}$  and the deviation of the relative inverse group velocity from some mean value $v^{-1}$. We thus have $\Delta k_0(z)=\bar{\Delta k}_0+\epsilon(z)$ and $\Delta v_g(z)^{-1}=v^{-1}-\epsilon(z)/(\Omega_3-\Omega_2)$. As a result, the $\epsilon(z)$ terms  cancel to give a constant detector energy gap $\tilde{\Omega}=\bar{\Delta k}_0-v^{-1}(\Omega_3-\Omega_2)$. The second assumption is that  the nonlinear susceptibility $\chi^{(2)}(z)$ can be shaped to compensate for variation in the refractive indices such that $\tilde{\eta}(z)=\tilde\eta$ is a constant. 

As an example, we consider Type-I KTP ($z\rightarrow y+y$), pumped by a quasi-monochromatic laser with frequency $\Omega_3=3.6\times 10^{15}$ rad/s ($\lambda\approx 523$ nm). To make the analogy, we require the UDW detector energy gap $\Omega$ and the phase-mismatch $\bar{\Delta k}_0$ to be positive. From the Sellmeier equations for KTP \cite{ktpsel}, we find this is satisfied when $\Omega_2=2\times 10^{15}$ rad/s ($\lambda\approx 942$ nm). This yields  a mean phase mismatch $\bar{\Delta k}_0=1.004\times10^{-9}$ m$^{-1}$ and a mean relative inverse group velocity $v^{-1}\approx 1.79\times 10^6$ m/s, which corresponds  to a detector energy gap $\Omega=\tilde\Omega v=\bar{\Delta k}_0v+\Omega_2-\Omega_3\approx 1.8\times 10^{14}$ rad/s. Since the crystal is not phasematched, it will not generate any photon pairs  in the absence of a relative-inverse-group-velocity gradient.

We take $\Delta v_g(z)^{-1}=\exp(-a  z/v)/v$, which gives $\tilde q(z)=\exp(-az/v  )/a$ (up to a phase factor on the amplitude). This corresponds to a uniformly accelerating detector, for which the amplitude is
\begin{align}
   \mathcal{A}'_\textsc{s} ={}&i\tilde\kappa\tilde\eta\int_{z_i}^{z_f}dz e^{i\tilde\Omega z}e^{-i\frac{\Omega_3-\Omega_2}{a} e^{-a z/v }} \,.
\end{align}
For a crystal of finite length, the integral above must be integrated numerically. This poses a  challenge because the integrand oscillates with a frequency that changes exponentially over the length of the crystal, limiting the crystal length over which we can make predictions.  For our physical parameters and integration approach, that limit is $\sim 100~\mu$m. We note that this does not restrict the experiment to those distances, simply our ability to model it. Longer crystals, however, will require a more drastic relative-inverse-group-velocity gradient $\Delta v_g(z)^{-1}$, which likely \emph{will} impose limits on the experiment. 

\begin{figure}
    \centering
    \includegraphics[width=\columnwidth]{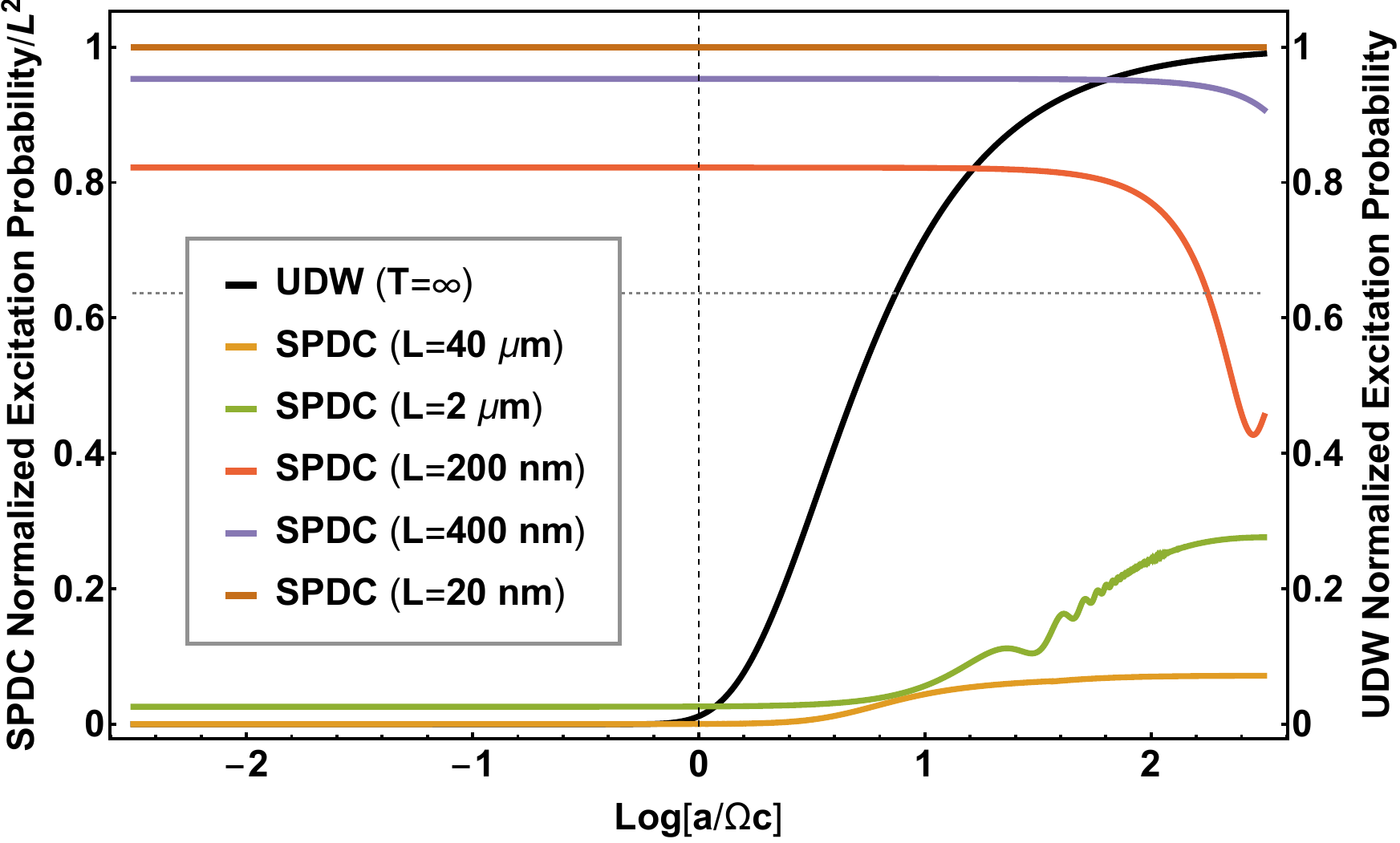}
    \caption{ Normalized detector excitation probabilities. For the UDW detector (black line), we plot $2\pi\Omega /(a(\exp(2\pi \Omega /a)-1)$, which has a Planckian form in $\Omega$ with a temperature $\beta^{-1}=a/2\pi$. For the SPDC system (coloured lines), we plot $|\mathcal{A}'_{\textsc{s}}/\tilde\kappa\tilde\eta L|^2$.  For longer crystals, the curve is qualitatively similar to the Planckian curve while for shorter crystals,  the excitation probability decreases with acceleration (this was also observed in \cite{brenna2016anti}, and may be related to the anti-Unruh effect \cite{garay2016thermalization}). For comparison, we plot $2/\pi$ (horizontal dashed line), which is the normalized excitation probability/$L^2$ for a standard periodically-poled crystal (periodicity $\Lambda\approx 3.5~\mu$m)  of length $L$ phasematched at $\omega_1=\Omega_1$ and $\omega_2=\Omega_2$.}
    \label{fig:fig2}
\end{figure}

We numerically evaluate the spectrum for the SPDC system as a function of the effective acceleration $a$ for various crystal lengths, and compare it with the known result for a UDW detector undergoing constant acceleration from $\tau_i=-\infty$ to $\tau_f=\infty$. We plot the corresponding excitation probabilities in FIG. \ref{fig:fig2}. Longer crystals have qualitatively similar behaviour to the UDW detector, where the excitation probability grows with acceleration. An interesting effect occurs for short crystals, however, where the excitation probability decreases with acceleration. This was also observed in \cite{brenna2016anti,garay2016thermalization}, and is likely related to the weak anti-Unruh effect.

Group velocity dispersion (GVD) can be engineered, e.g., by varying the cross-sectional shape and area of the waveguide \cite{ahmad2018modeling} or by varying the distance between coupled cavities in a photonic crystal slab \cite{fussell2007engineering}. There may exist a material or parameter regime where realistic GVD variations will lead to reasonable count rates, but finding these will require an extensive parameter search (made complicated by the need to integrate highly-oscillating functions), as well as extension of the analysis to three-dimensional waveguides. We leave this for future work.  

In the mean-time, the analogy between the acceleration of a UDW detector and the relative dispersion raises interesting questions. Is it possible to use GVD engineering as a new approach to phase matching? On the other hand, how should one think about the analogue of periodic poling---a common quasi-phasematching technique in NLO--- in the context of a UDW detector? And how does the analogy introduced in this paper relate to the analogy between the refractive index of a dielectric and curvature of spacetime \cite{novello2002artificial}? These would be interesting to explore further.

The Unruh effect lies at the intersection of thermal physics, quantum physics and gravity. It's an important signpost in the search for the theory of quantum gravity \cite{earman2011unruh,smolin2008three}. Unfortunately, the effect is yet to be verified experimentally. While simulations of some Unruh-related phenomena have been studied in various systems, including NLO, this is the first proposal for simulating an Unruh-DeWitt detector \emph{using optical modes and a nonlinear material}. We expect this analogy to be fruitful in the cross-pollination of ideas among different areas of physics. 

\section{Acknowledgements}
AMB thanks Eduardo Mart\'in-Mart\'inez and Robb Mann for helpful discussions. Research at Perimeter Institute is supported by the Government of  Canada through Industry Canada and by the Province of Ontario through the Ministry of Research and Innovation. This research was supported in part by the Natural Sciences and Engineering Research Council of Canada (RGPIN-2016-04135), Canada Research Chairs, Industry Canada and the Transformative Quantum Technologies program.

\onecolumngrid

\appendix

\section{The Unruh-DeWitt transition amplitude}\label{sec:UDWA}
We now compute $ \bra{\omega}\phi(x(\tau)t(\tau))\ket{0}$, where $\ket{\omega}=a\dg(\omega)\ket{0}$: 
\begin{align}
 \bra{\omega}\phi(x(\tau)t(\tau))\ket{0}={}&\bra{0}a\dg(\omega)\left(\int \frac{ d\omega_{k}}{\sqrt{4\pi\omega_{k}}} (a(\omega_{k}) e^{-i\omega_{k}q(\tau)} + a\dg(\omega_{k}) e^{i\omega_{k}q(\tau)})\right)\ket{0}\\
 ={}&\int \frac{ d\omega_{k}}{\sqrt{4\pi\omega_{k}}} ( e^{-i\omega_{k}q(\tau)}\bra{0}a\dg(\omega)a(\omega_{k})\ket{0} +  e^{i\omega_{k}q(\tau)}\bra{0}a\dg(\omega)a\dg(\omega_{k})\ket{0})\\
  ={}&\int \frac{ d\omega_{k}}{\sqrt{4\pi\omega_{k}}}  e^{-i\omega_{k}q(\tau)}\delta(\omega-\omega_k)\\
  ={}&\frac{ e^{-i\omega q(\tau)}}{{\sqrt{4\pi\omega}} }\,.
\end{align}

Therefore the amplitude for transition from the ground state $|g, 0\rangle$ to the state $|e, \omega \rangle $ (where $g$ and $e$ represent the ground and excited states of the detector respectively; and  $0$ and $k$ represent the  ground and single-photon-excited states of the field) at first order expansion of the evolution operator is:
\begin{align}
   \mathcal{A}_{\textsc{u},\omega}(\Omega) ={}&-i \bra{e,\omega}\int_{\tau_i}^{\tau_f} \dd \tau H_{\textsc{u}}(\tau)\ket{g,0}\\
   ={}&-i \bra{e,\omega}\int_{\tau_i}^{\tau_f} \dd \tau \lambda\eta(\tau) m(\tau)\phi(x(\tau),t(\tau))\ket{g,0}\\
    ={}&-i \int_{\tau_i}^{\tau_f} \dd \tau \lambda\eta(\tau) \bra{e}m(\tau)\ket{g}\bra{\omega}\phi(x(\tau),t(\tau))\ket{0}\\
    ={}&-\frac{i\lambda\mathcal{M}}{\sqrt{4\pi  \omega_{k}}} \int_{\tau_i}^{\tau_f} \dd \tau \eta(\tau)  e^{-i\omega q(\tau)}\,,
\end{align}
with $\mathcal{M} = \langle e|m(\tau) | g \rangle=\bra{e}e^{iH_0\tau}m(0)e^{-iH_0\tau}\ket{g}=\bra{e}m(0)\ket{g}e^{i\Omega\tau}$. This can then be written as 
\begin{align}
   \mathcal{A}_{\textsc{u},\omega}(\Omega)    ={}&b(\omega_{k}) \int_{\tau_i}^{\tau_f} \dd \tau \eta(\tau)  e^{i\Omega\tau}e^{-i\omega q(\tau)}\,,
\end{align}
where $ b(\omega)=-i\lambda /\sqrt{4\pi\omega} \bra{e}m(0)\ket{g}$.

\section{The SPDC transition amplitude}\label{sec:spdcder}

The interaction Hamiltonian for a three-wave mixing process in a waveguide is given in \cite{yang2008spontaneous} as: 
\begin{align}\label{eq:H2}
\begin{split}
H_{_\textsc{s}}(t)={}&\frac{1}{3\epsilon_0}\int d\mathbf{r}\Gamma^{ijk}_{(2)}(\mathbf{r})\hat{D}^{i}(\mathbf{r},t)\hat{D}^{j}(\mathbf{r},t)\hat{D}^{k}(\mathbf{r},t)\,,
\end{split}
\end{align}
where $\Gamma^{ijk}_{(2)}(\mathbf{r})$ is a tensor related to the more commonly-used nonlinear tensor $\chi^{ijk}_{(2)}(\mathbf{r})$ via Eq. (15) in \cite{yang2008spontaneous}, and $D^{i/j/k}$ are the quantized displacement field operators. Under the assumptions that the tensor nature of $\chi_{(2)}(\mathbf{r})$ and the vector nature of the mode profiles of the displacement field can be neglected, $\chi_{(2)}$ can be taken to characterize the strength of the nonlinearity. Let us assume that we have three fields present in the process (labelled by $j=1,2,3$), which all have zero overlap with each other, either due to non-overlapping frequencies or orthogonal polarization. Let us also assume that the fields propagate collinearly along $z$ through a waveguide of length $L$ with only one transverse mode that is uniform over a cross-sectional area $A$.  Furthermore, let us assume that one of the fields ($j=3$) is intense, non-depleting, and can be treated classically. The interaction Hamiltonian can then be written as (see Eq. (3.28) of \cite{quesada2015very}):
\begin{align}\label{eq:HNx}
\begin{split}
  H_{_\textsc{s}}(t)={}&  -\hbar \epsilon\int_{0}^{\infty}d\omega_1\int_{0}^{\infty} d\omega_2 \int_{0}^{\infty}d\omega_3~e^{i(\omega_1+\omega_2-\omega_3)t}~\mathrm{sinc}\left(\frac{L}{2}\left(k_1(\omega_1)+k_2(\omega_2)-k_3(\omega_3)\right)\right)\\
  &\times\alpha(\omega_3)\hat{a}\dg_{1}(\omega_1)\hat{a}\dg_{2}(\omega_2) +\hc
  \end{split}
\end{align}
where $k_j(\omega)=\omega n_j(\omega)/c$, where $n_j(\omega)$ is the refractive index of mode $j$, where
\begin{align}
    \epsilon=2L\chi^{(2)}\sqrt{\frac{\sqrt{2}U_0\pi\Omega_1\Omega_2}{\sqrt{\pi}(4\pi)^3\epsilon_0Ac^3n_1(\Omega_1)n_2(\Omega_2)n_3(\Omega_3)\tau}}\,,
\end{align}
where  $\Omega_j$ are the central frequencies of the fields, $U_0$ is the energy of the classical pulse  used to pump the crystal and 
\begin{align}\label{eq:pulse}
    \alpha(\omega_3)=\frac{\tau}{\sqrt{\pi}}e^{-\tau^2(\omega_3-\Omega_3)^2}
\end{align}
is the shape of the pulse, which has units of time, where $\tau=1/\sqrt{\sigma}=2\sqrt{2\ln 2} \quad\mathrm{FWHM}$, where $\sigma$ is the standard deviation of the Gaussian distribution in frequency, and FWHM is the full width at half maximum of a Gaussian distribution. 

We now generalize this result to include a spatial dependence on the nonlinear susceptibility. It is not necessary to repeat the derivation from \cite{quesada2015very} from first principles for the general case. We can just reverse some of the final steps in that derivation and make our generalizations at the appropriate points. We first note that 
\begin{align}
   \chi^{(2)}\mathrm{sinc}\left(\frac{L}{2}\left(k_1(\omega_1)+k_2(\omega_2)-k_3(\omega_3)\right)\right)= \frac{1}{L}\int_{z_i}^{z_f}dz  \chi^{(2)} e^{-i(k_1(\omega_1)+k_2(\omega_2)-k_3(\omega_3))z}\,,
\end{align}
where $z_i=-L/2$ and $z_f=L/2$, and then make the substitution $\chi^{(2)}\rightarrow \chi^{(2)}(z)$.  We also include a spatial dependence on the refractive index by making the substitution $n_j(\omega)\rightarrow n_j(\omega;z)$, which in turn requires the substitution
\begin{align}
    (k_1(\omega_1)+k_2(\omega_2)-k_3(\omega_3))z\rightarrow \int_{z_i}^{z}d\zeta(k_1(\omega_1;\zeta)+k_2(\omega_2;\zeta)-k_3(\omega_3;\zeta))\,,
\end{align}
 where the integral over $\zeta$ keeps track of the phase picked up as the light propagates through a material with variable $k_j$  \cite{helmfrid1993influence,santandrea2019fabrication}.

The interaction Hamiltonian becomes
\begin{align}\label{eq:Hz}
\begin{split}
  H_{_\textsc{s}}(t)={}&  -\frac{\hbar \kappa}{2\pi}\int_{0}^{\infty}d\omega_1\int_{0}^{\infty} d\omega_2 \int_{0}^{\infty}d\omega_3~e^{i(\omega_1+\omega_2-\omega_3)t} \alpha(\omega_3)\hat{a}\dg_{1}(\omega_1)\hat{a}\dg_{2}(\omega_2) \\
  &\times \int_{z_i}^{z_f}dz  \frac{\chi^{(2)}(z)}{\sqrt{n_1(\Omega_1;z)n_2(\Omega_2;z)n_3(\Omega_3;z)}} e^{-i\int_{-L/2}^{z}d\zeta(k_1(\omega_1;\zeta)+k_2(\omega_2;\zeta)-k_3(\omega_3;\zeta))}+\hc
  \end{split}
\end{align}
where
\begin{align}\label{eq:kappa}
    \kappa=4\pi\sqrt{\frac{\sqrt{2}U_0\pi\Omega_1\Omega_2}{\sqrt{\pi}(4\pi)^3\epsilon_0Ac^3\tau}}\,.
\end{align}

We now use this Hamiltonian to derive the transition amplitude in Eq. \eqref{eq:AUDx}. We start with the definition in Eq. \eqref{eq:AUDx0} and insert the expression for the Hamiltonian in Eq. \eqref{eq:Hz}, to give
\begin{align}
\begin{split}
   \mathcal{A}_\textsc{s} ={}&i\frac{\kappa}{2\pi} \int_{-\infty}^{\infty} \dd t    \int_{0}^{\infty}d\omega_3~e^{i(\omega+\Omega_2-\omega_3)t} \alpha(\omega_3) \\
  &\times \int_{z_i}^{z_f}dz  \frac{\chi^{(2)}(z)}{\sqrt{n_1(\Omega_1;z)n_2(\Omega_2;z)n_3(\Omega_3;z)}} e^{-i\int_{-L/2}^{z}d\zeta(k_1(\omega;\zeta)+k_2(\Omega_2;\zeta)-k_3(\omega_3;\zeta))}\,.
  \end{split}
\end{align}
If we now perform the integral over time, we get
\begin{align}
\begin{split}
   \mathcal{A}_\textsc{s} ={}&i\kappa    \int_{0}^{\infty}d\omega_3~\delta(\omega+\Omega_2-\omega_3) \alpha(\omega_3) \\
  &\times \int_{z_i}^{z_f}dz  \frac{\chi^{(2)}(z)}{\sqrt{n_1(\Omega_1;z)n_2(\Omega_2;z)n_3(\Omega_3;z)}} e^{-i\int_{-L/2}^{z}d\zeta(k_1(\omega;\zeta)+k_2(\Omega_2;\zeta)-k_3(\omega_3;\zeta))}\,,
  \end{split}
\end{align}
and perform the integral over $\omega_3$ to give
\begin{align}
   \mathcal{A}_\textsc{s} ={}&i\kappa     \alpha(\Omega_2+\omega)  \int_{z_i}^{z_f}dz  \frac{\chi^{(2)}(z)}{\sqrt{n_1(\Omega_1;z)n_2(\Omega_2;z)n_3(\Omega_3;z)}} e^{-i\int_{-L/2}^{z}d\zeta\Delta k(\Omega_2,\omega;\zeta)}\,,
\end{align}
where $\Delta k(\Omega_2,\omega;\zeta)=k_1(\omega;\zeta)+k_2(\Omega_2;\zeta)-k_3(\Omega_2+\omega;\zeta)$.

The pump laser is centred at frequency $\Omega_3$. We can then take $\alpha(\Omega_2+\omega)$ to be a function of $\omega$ centred about $\omega=\Omega_1=\Omega_3-\Omega_2$, and assume that $\Delta k(\Omega_2,\omega;\zeta)$ is close to linear in $\omega$ where   $\alpha(\Omega_2+\omega)$ is non-negligible.  We then do a first-order expansion of $\Delta k(\Omega_2,\omega;\zeta)$ about $\omega=\Omega_1$.  The terms are
\begin{subequations}
\begin{align}
    k_1(\omega;z)={}&k_1(\Omega_1;z)+\frac{1}{v_1(z)}(\omega-\Omega_1)\\
    k_2(\Omega_2;z)={}&k_2(\Omega_2;z)\\
    k_3(\Omega_2+\omega;z)={}&k_3(\Omega_3;z)+\frac{1}{v_3(z)}(\omega-\Omega_1)\,.
\end{align}
\end{subequations}
where $v_1(z)=(\partial k_1(\omega;z)/\partial\omega|_{\omega=\Omega_1})^{-1}$ and $v_3(z)=(\partial k_3(\Omega_2+\omega;z)/\partial\omega|_{\omega=\Omega_1})^{-1}$ are the group velocities for modes 1 and 3. Using $\Omega_1=\Omega_3-\Omega_2$, we can write $\Delta k(\Omega_2,\omega;z)$  as 
\begin{align}
    \Delta k(\Omega_2,\omega;z)=\Delta k_0(z)- \frac{1}{\Delta v_g(z)}(\Omega_3-\Omega_2)+ \frac{1}{\Delta v_g(z)}\omega\,,
\end{align}
where 
\begin{align}\label{eq:dk00}
    \Delta k_0(z)={}&k_3(\Omega_3;z)-k_2(\Omega_2;z)-k_1(\Omega_1;z)
\end{align}
is the phase mismatch, and
\begin{align}\label{eq:dk10}
   \frac{1}{\Delta v_g(z)}={}&\frac{1}{v_3(z)}-\frac{1}{v_1(z)}
\end{align}
is the relative inverse group velocity. 
We now identify 
\begin{align}\label{eq:Omega2}
    \Delta k(\Omega_2,\omega_s;z)=\underbrace{\Delta k_0(z)-\frac{1}{\Delta v_g(z)}(\Omega_3-\Omega_2)}_{\tilde\Omega(z)}+ \underbrace{\frac{1}{\Delta v_g(z)}}_{\frac{d}{dz}\tilde q(z)}\omega\,.
\end{align}
 With all of this, we can then write
\begin{align}
   \mathcal{A}_\textsc{s} ={}& b_\textsc{s}(\omega)\int_{z_i}^{z_f}dz\tilde\eta(z) e^{i\tilde\Omega(z) z}e^{i\omega\tilde q(z)}\,,
\end{align}

where 
\begin{subequations}
\begin{align}
   b_\textsc{s}(\omega)={}&i\kappa     \alpha(\Omega_2+\omega) \\
    \tilde \Omega(z)={}&\Delta k_0(z)-\frac{1}{\Delta v_g(z)}(\Omega_3-\Omega_2)\\
    \tilde\eta(z)={}&\frac{\chi^{(2)}(z)}{\sqrt{n_1(\Omega_2+\Omega_3;z)n_2(\Omega_2;z)n_3(\Omega_3;z)}} \\
    \tilde q(z)={}&\int_{z_i}^{z}\frac{1}{\Delta v_g(\zeta)} d\zeta\,.
\end{align}
\end{subequations}

\end{document}